arXiv: 0802.3424

# Property of Tsallis entropy and principle of entropy increase

Du Jiulin

*Department of Physics, School of Science, Tianjin University, Tianjin 300072, China*

E-mail: jiulindu@yahoo.com.cn

**Abstract** The property of Tsallis entropy is examined when considering tow systems with different temperatures to be in contact with each other and to reach the thermal equilibrium. It is verified that the total Tsallis entropy of the two systems cannot decrease after the contact of the systems. We derived an inequality for the change of Tsallis entropy in such an example, which leads to a generalization of the principle of entropy increase in the framework of nonextensive statistical mechanics.

## 1. Introduction

In recent years, a generalization of Bltzmann-Gibbs(B-G) statistical mechanics initiated by Tsallis has focused a great deal of attention, the results from the assumption of nonadditive statistical entropies and the maximum statistical entropy principle, which has been known as "Tsallis statistics" or nonextensive statistical mechanics(NSM) (Abe and Okamoto, 2001). This generalization of B-G statistics was proposed firstly by introducing the mathematical expression of Tsallis entropy (Tsallis, 1988) as follows,

$$S_q = \frac{k}{1-q}(\int \rho^q d\Omega - 1) \tag{1}$$

where $k$ is the Boltzmann's constant. For a classical Hamiltonian system, $\rho$ is the phase space probability distribution of the system under consideration that satisfies the normalization $\int \rho \, d\Omega = 1$ and $d\Omega$ stands for the phase space volume element. The entropy index $q$ is a positive parameter whose deviation from unity is thought to describe the degree of nonextensivity. The conventional B-G entropy $S_B$ is obtained as the limit $q \to 1$,

$$S_B = \lim_{q \to 1} S_q = -k \int \rho \ln \rho \, d\Omega \tag{2}$$

One distinctive difference between the Tsallis entropy $S_q$ and the B-G entropy $S_B$ lies in its pseudo-additive property. If $a$ and $b$ are two independent systems in the sense that the probability distribution of $a+b$ factorizes into those of $a$ and of $b$, then Tsallis entropy of the composite system $a+b$ can be expressed by

$$S_q(a+b) = S_q(a) + S_q(b) + (1-q)S_q(a)S_q(b)/k \tag{3}$$

Such a generalization may address the nonextensive systems with long-range interactions, long-range microscopic memory, and fractal or multifractal relevant space-time. It retains some of the structure of the standard theory such as the Legendre transform structure, $H$ theorem, Onsager reciprocity theorem, fluctuation-dissipation theorem, zeroth law of thermodynamics, equipartition and virial theorem etc (please see

http://tsallis.cat.cbpf.br/biblio.htm). Some applications of NSM in the astrophysical self-gravitating systems(Du 2006,2007 and references therein) and in the plasma systems (Du 2004 and the references therein; Lavagno & Quarati 2006; Shaikh et al. 2007) with the long-range interactions have been investigated.

When the traditional B-G statistical mechanics is generalized, it is most important to establish consistency of the new theory with the basic principles of thermodynamics. In this paper, we study the principle of entropy increase by using Tsallis entropy in the NSM and try to give a mathematical formulation about the second law of thermodynamics in the framework of NSM.

## 2. Entropy increase in B-G statistics

In B-G statistical mechanics, the principle of entropy increase is known a mathematical formulation of the second law of thermodynamics. It states that when a system undergoes an adiabatic process its entropy can never be decreased, the entropy is unchanged if the process is reversible, or the entropy increases if the process is irreversible. The principle of entropy increase can be verified from the canonical ensemble theory by studying changes of the entropy if tow systems with different temperatures are in contact with each other and become in thermal equilibrium.

We consider tow extensive systems, their equilibrium properties are determined by the canonical probability distribution $\rho_1$ and $\rho_2$, and the corresponding phase space volume elements are denoted by $d\Omega_1$ and $d\Omega_2$, respectively. Without the contact, the tow systems are independent and are regarded as one system. The probability distribution for the total system and phase space volume element are $\rho_0 = \rho_1 \rho_2$ and $d\Omega = d\Omega_1 d\Omega_2$, respectively, and the entropy is considered in terms of the additivity as

$$S_{B0} = -k\int \rho_1 \ln \rho_1 d\Omega_1 - k\int \rho_2 \ln \rho_2 d\Omega_2 = -k\int \rho_0 \ln \rho_0 d\Omega \qquad (4)$$

with $\rho_1 = \frac{1}{Z_1} e^{-\beta_1 H_1}$ and $\rho_2 = \frac{1}{Z_2} e^{-\beta_2 H_2}$. The energy of every subsystem is expressed, respectively, by

$$<H_1> = \int \rho_1 H_1 d\Omega_1 = \int \rho_0 H_1 d\Omega \,; \quad <H_2> = \int \rho_0 H_2 d\Omega \qquad (5)$$

After the contact, the new equilibrium in the total system is determined by the canonical probability distribution $\rho = \frac{1}{Z}e^{-\beta H}$. The corresponding entropy is given as

$$S_B = -k\int \rho \ln \rho \, d\Omega. \qquad (6)$$

Then, the change of B-G entropy can be calculated by

$$\begin{aligned} S_B - S_{B0} &= k\int (\rho_0 \ln\rho_0 - \rho \ln \rho)\, d\Omega \\ &= k<X(\rho_0,\rho)>_0 - k\int (\rho-\rho_0)\ln \rho \, d\Omega \\ &= k<X(\rho_0,\rho)>_0 - k\beta \int (\rho-\rho_0) H \, d\Omega \end{aligned} \qquad (7)$$

where $X(\rho_0,\rho) = \ln \rho_0 - \ln \rho$ and $<X(\rho_0,\rho)>_0 = \int \rho_0 X(\rho_0,\rho)d\Omega$. It has been verified (Wang Zhuxi, 1965) that the inequality $<X(\rho_0,\rho)>_0 \geq 0$ is always satisfied and the equality holds if and only if we have $\rho = \rho_0$. Suppose the interaction between the tow subsystems is small so that we can write $H = H_1 + H_2$ and, after the contact, the total energy is unchanged, then the second term on the right hand side of Eq.(7) vanishes, we have

$$S_B - S_{B0} \geq 0 \qquad (8)$$

The tow subsystems being in contact with each other can be regarded as an isolated system, in which any processes taking place can be treated as the adiabatic ones. This inequality shows that the entropy cannot be decreased in the adiabatic processes. The entropy is unchanged only if $\rho = \rho_0$, which holds only if the tow systems have the same temperature: $\beta_1 = \beta_2 = \beta$. The inequality (8) can be thought as the mathematical expression of the principle of entropy increase in the above example.

**3. Principle of entropy increase for the nonextensive system**

Now we deal with the property of Tsallis entropy for the nonextensive systems in NSM and try to give a generalization of the principle of entropy increase, Eq.(8). We consider two nonextensive systems with different temperatures and with the Hamiltonians $H_1$ and $H_2$, respectively. Their Tsallis' entropies are given, respectively, by

$$S_{q1} = \frac{k}{1-q}\left(\int \rho_1^q d\Omega_1 - 1\right); \quad S_{q2} = \frac{k}{1-q}\left(\int \rho_2^q d\Omega_2 - 1\right) \tag{9}$$

In NSM, the corresponding optimal probability distributions and the generalized canonical partition functions are found (Tsallis,1998), respectively, to be

$$\rho_1 = \frac{1}{Z_{q1}}[1-(1-q)\beta_1^*(H_1 - U_{q1})]^{\frac{1}{1-q}}; \tag{10}$$

$$\rho_2 = \frac{1}{Z_{q2}}[1-(1-q)\beta_2^*(H_2 - U_{q2})]^{\frac{1}{1-q}} \tag{11}$$

$$Z_{q1} = \int [1-(1-q)\beta_1^*(H_1 - U_{q1})]^{\frac{1}{1-q}} d\Omega_1 \tag{12}$$

$$Z_{q2} = \int [1-(1-q)\beta_2^*(H_2 - U_{q2})]^{\frac{1}{1-q}} d\Omega_2 \tag{13}$$

where $U_{q1}$ and $U_{q2}$ are given in terms of the constraints on the generalized internal energy $<H_1>_q$ and $<H_2>_q$, defined by the normalized $q$-expectation values,

$$U_{q1} = <H_1>_q = \int \rho_1^q H_1 d\Omega_1 / \int \rho_1^q d\Omega_1, \tag{14}$$

$$U_{q2} = <H_2>_q = \int \rho_2^q H_2 d\Omega_2 / \int \rho_2^q d\Omega_2. \tag{15}$$

$\beta_1^* = \beta_1 / \int \rho_1^q d\Omega_1$ and $\beta_2^* = \beta_2 / \int \rho_2^q d\Omega_2$ with $\beta_1$ and $\beta_2$ the Lagrange multipliers associated with the tow energy constraints above, respectively. $\beta_1$ and $\beta_2$ are identified with their inverse temperatures.

Without the contact, these tow nonextensive systems are independent. The probability distribution function and the phase space volume element of the total system are $\rho_0 = \rho_1 \rho_2$ and $d\Omega = d\Omega_1 d\Omega_2$, respectively. The total energy is

$$U_{q1} + U_{q2} = \int \rho_0^q (H_1 + H_2) \, d\Omega / \int \rho_0^q d\Omega \tag{16}$$

Tsallis' entropy of the total system satisfies the pseudoadditivity of Eq.(3), namely

$$S_{q0} = S_{q1} + S_{q2} + \frac{1-q}{k} S_{q1} S_{q2} = \frac{k}{1-q}\left(\int \rho_0^q d\Omega - 1\right) \tag{17}$$

After the contact and being in thermal equilibrium, new Tsallis' entropy $S_q$ of the total system is expressed by Eq.(1) with the optimal probability distribution and the generalized canonical partition function, respectively,

$$\rho = \frac{1}{Z_q}[1-(1-q)\beta^*(H-U_q)]^{\frac{1}{1-q}}; \quad (18)$$

$$Z_q = \int[1-(1-q)\beta^*(H-U_q)]^{\frac{1}{1-q}}d\Omega \quad (19)$$

where $H$ is the new Hamiltonian of the total system, $U_q$ is the normalized $q$-expectation value of the total energy, $U_q = <H>_q = \int \rho^q H \, d\Omega / \int \rho^q d\Omega$, and $\beta^* = \beta / \int \rho^q d\Omega$ with $\beta$ the Lagrange multiplier associated with this energy constraint and identified with the inverse temperature $T$ of the total system. Then, the change of the entropy after and before the contact of the tow nonextensive systems can be calculated as

$$S_q - S_{q0} = \frac{k}{1-q}\int(\rho^q - \rho_0^q)d\Omega = \frac{k}{1-q}\int \rho^q\left[1-\left(\frac{\rho_0}{\rho}\right)^q\right]d\Omega \quad (20)$$

Let $X_q(\rho_0,\rho) = k[1-(\rho_0/\rho)^q]/(1-q)$ and $c_q = \int \rho^q d\Omega$, Eq.(20) becomes

$$S_q - S_{q0} = c_q <X_q(\rho_0,\rho)>_q \quad (21)$$

where we have introduced the so-called $q$-expectation value of $X_q(\rho_0,\rho)$ by $<X_q>_q = \int \rho^q X_q d\Omega / \int \rho^q d\Omega$. It is further written as

$$<X_q>_q = \frac{1}{c_q}\frac{k}{1-q}\int \rho^q[1-(\rho_0/\rho)^q]d\Omega =$$

$$= \frac{k}{1-q}\int \rho[1-(\rho_0/\rho)^q]d\Omega - \frac{1}{c_q}\frac{k}{1-q}\int(\rho^q - \rho_0^q)(c_q\rho^{1-q}-1)d\Omega$$

$$= <X_q> + \frac{k\beta^*}{c_q}\int(\rho^q - \rho_0^q)(H-U_q)d\Omega \quad (22)$$

where $<X_q> = \int \rho X_q d\Omega$ is the standard expectation value of $X_q(\rho_0,\rho)$. With $(\rho_0/\rho) > 0$, $q>0$, we find the inequality,

$$X_q = \frac{k}{1-q}\left[1-\left(\frac{\rho_0}{\rho}\right)^q\right] \geq \frac{qk}{1-q}\left[1-\left(\frac{\rho_0}{\rho}\right)\right], \quad (23)$$

and the equality holds if and only if $\rho = \rho_0$. Consequently, we have, for $q>0$,

$$<X_q> \geq 0 \tag{24}$$

If suppose the interaction between the tow nonextensive subsystems is small so that, after the contact, the Hamiltonian of total system can be still written as $H = H_1 + H_2$ and the total energy remains unchanged, namely

$$\frac{\int \rho^q H \, d\Omega}{\int \rho^q d\Omega} = \frac{\int \rho_0^q H \, d\Omega}{\int \rho_0^q d\Omega} \tag{25}$$

then the second term on the right hand side of Eq.(22) vanishes. From Eq.(21) we find that the inequality for Tsallis entropy becomes

$$S_q - S_{q0} \geq 0 \tag{26}$$

This inequality shows that, after the tow nonextensive systems are in contact with each other, the total Tsallis entropy cannot be decreased, but it is unchanged if and only if $\rho = \rho_0$. The tow systems being in contact with each other can be treated as an isolated system, in which any processes taking place inside the system can be regarded as adiabatic ones. Therefore, we have verified the principle of entropy increase for Tsallis entropy in the above example of the nonextensive systems and have obtained the generalization of the inequality (8) in the framework of NSM.

## 4. Conclusion

In conclusion, we have studied the property of Tsallis entropy and have examined the principle of entropy increase in the framework of NSM. By considering tow nonextensive systems with different temperatures are in contact with each other and are in thermal equilibrium, we investigate the change of Tsallis' entropy and we demonstrate that, after the contact, the Tsallis entropy of the total nonextensive system cannot be decreased, which leads to a generalization of the principle of entropy increase in the framework of NSM.

*Additional remarks*: The definition of heat in NSM was suggested by Rajagopal (2003) with density matrix in the quantum theory and a similar formulation to the Clausius' inequality was established by Abe and Rajagopal (2003) in the framework of nonextensive quantum thermodynamics.

**Acknowledgment**

I hope to thank Professor H. J. Haubold and the Second UN/NASA Workshop on the International Heliophysical Year and Basic Space Science (27 Nov.-1 Dec. 2006, Bangalore, India) for the invitation. This work is supported by the project of 985 Program of TJU in China and by the National Natural Science Foundation of China under grant No.10675088.